\documentclass[12pt]{article}          %%
\usepackage{graphicx}
\makeatletter  

\@addtoreset{equation}{section} \makeatother

\begin{document}

\renewcommand{\thefootnote}{\arabic{footnote}}

\newcommand{\hs}{\hspace*{0.5cm}}
\newcommand{\vs}{\vspace*{0.5cm}}
\newcommand{\be}{\begin{equation}}
\newcommand{\ee}{\end{equation}}
\newcommand{\bea}{\begin{eqnarray}}
\newcommand{\eea}{\end{eqnarray}}
\newcommand{\bary}{\begin{array}}
\newcommand{\eary}{\end{array}}
\newcommand{\bit}{\begin{itemize}}
\newcommand{\eit}{\end{itemize}}
\newcommand{\ben}{\begin{enumerate}}
\newcommand{\een}{\end{enumerate}}
\newcommand{\crn}{\nonumber \\}
\newcommand{\noi}{\noindent}
\newcommand{\al}{\alpha}
\newcommand{\la}{\lambda}
\newcommand{\bet}{\beta}
\newcommand{\ga}{\gamma}
\newcommand{\va}{\varphi}
\newcommand{\ci}[1]{\cite{#1}}
\newcommand{\re}[1]{(\ref{#1})}
\newcommand{\bi}[1]{\bibitem{#1}}
\newcommand{\lab}[1]{\label{#1}}
\newcommand{\ra}{\rightarrow}
\newcommand{\om}{\omega}
\newcommand{\pa}{\partial}
\newcommand{\fr}{\frac}
\newcommand{\sm}{\sigma}
\newcommand{\bc}{\begin{center}}
\newcommand{\ec}{\end{center}}
\newcommand{\nn}{\nonumber}
\newcommand{\Ga}{\Gamma}
\newcommand{\de}{\delta}
\newcommand{\De}{\Delta}
\newcommand{\ep}{\epsilon}
\newcommand{\varep}{\varepsilon}
\newcommand{\vthe}{\vartheta}
\newcommand{\ka}{\kappa}
\newcommand{\La}{\Lambda}
\newcommand{\vr}{\varrho}
\newcommand{\si}{\sigma}
\newcommand{\Si}{\Sigma}
\newcommand{\ta}{\tau}
\newcommand{\up}{\upsilon}
\newcommand{\Up}{\Upsilon}
\newcommand{\kh}{\chi}
\newcommand{\ze}{\zeta}
\newcommand{\ps}{\psi}
\newcommand{\Ps}{\Psi}
\newcommand{\ph}{\phi}
\newcommand{\vph}{\varphi}
\newcommand{\Ph}{\Phi}
\newcommand{\Om}{\Omega}

\def\lappeq{\mathrel{\rlap{\raise.5ex\hbox{$<$}}
{\lower.5ex\hbox{$\sim$}}}}

\bc {\huge  Michel  parameter  in  3-3-1 model \\
with three lepton singlets}

\vspace*{0.5cm}

{\bf Hoang Ngoc Long}

\vspace*{0.5cm}

{\it  Institute of  Physics, Vietnam Academy of Science and Technology, \\
10 Dao Tan, Ba Dinh, Hanoi, Vietnam}\\
\ec

\begin{abstract}
We show that the mass matrix of electrically neutral gauge bosons in the recently proposed
model based on $\mathrm{SU}(3)_C\otimes \mathrm{SU}(3)_L
\otimes \mathrm{U}(1)_X$ group with three lepton singlets \cite{331b}  has two exact
eigenvalues: a  zero  corresponding the photon  mass and the second one  equaling the mass
of the imaginary component $A_{5\mu}$.
Hence the neutral non-Hermitian gauge boson $X^0_\mu$ (defined as $\sqrt{2} X^0_\mu =
A'_{4\mu} - i A_{5\mu}$)  is properly determined.
With extra vacuum expectation value of the Higgs field $n_2$, there are mixings among the Standard Model
 $W$  boson and the extra charged   gauge boson $Y$ carrying lepton number 2 (bilepton)
    as well as among
neutral gauge bosons $Z, Z'$ and $X^0$. These mixings lead to very rich phenomenology of the model.
 The leading order of the Michel  parameter ($\rho$) has quite special form  requiring
an  equality of the vacuum expectation values in the second step of spontaneous
symmetry breaking, namely,  $k_1=k_2$.
\end{abstract}


\section{Introduction \label{introd}}

At present, it is well known that neutrinos are massive that contradicts the
Standard Model (SM).
The experimental data \cite{pdg} show that masses of neutrinos are tiny small and neutrinos mix with
special pattern in  approximately tribimaximal form \cite{hps}.
The neutrino masses, dark matter and the baryon asymmetry of Universe (BAU) are the facts requiring
extension of the SM.

Among the extensions beyond the SM, the models based on $\mathrm{SU}(3)_C\otimes \mathrm{SU}(3)_L \otimes \mathrm{U}(1)_X$ (3-3-1)
 gauge group  \cite{ppf, r331} have some interesting features including the ability to explain why there exist
 three families of quarks and leptons \cite{ppf, r331} and the electric charge quantization \cite{chargequan}. In this scheme the gauge couplings can be unified
at the scale of order TeV\emph{ without  supersymmetry} \cite{uni}.

 Concerning the content in lepton triplet, there exist two main versions of 3-3-1 models: the minimal version \cite{ppf} without extra lepton and the model with right-handed neutrinos \cite{r331} without exotic charged particles.
 Due to the fact that particles with different lepton numbers lie in the same triplet, the lepton number
 is violated and it is better to deal with a new conserved charge $\mathcal{L}$ commuting with the gauge symmetry \cite{clong}
 \be
 L=\frac{4}{\sqrt{3}} T_8+\mathcal{L}\, .
 \label{eq81}
 \ee
 In the framework of the 3-3-1 models, almost issues concerning neutrino physics are solvable. In  the minimal 3-3-1 model where perturbative regime is trustable until 4-5 TeV, to realize idea of seesaw, the effective dimension-5 operator is used \cite{pires}. In regard to the 3-3-1 model with right-handed neutrinos, effective-5 operators are sufficient to generate light neutrino masses. The effective dimension-5 operator may be realized through a kind of type-II seesaw mechanism implemented by a sextet of scalars belonging to the GUT scale \cite{donglong2008}. There are some ways  to explain smallness of neutrino  masses:  the radiative mechanism,  the seesaw one or their combination - radiative
 seesaw.  The seesaw mechanism is   the most easy and elegant way of generating small neutrino masses by using the Majorana neutrinos with mass belonging to  GUT scale.  With such high scale, the Majorana neutrinos are unavailable
  for laboratory searches. The existence of sextet is unfavorableness because of lack predictability  associated with it.
  There are attempts to improve the situation.

  In the recently proposed  model \cite{331b}, the authors have introduced three  lepton/neutrino singlets and used radiative mechanism to get a model, where the seesaw mechanism is realized at quite low scale of few TeVs.
  We remind that in the 3-3-1 model with right-handed
   neutrinos, there are two scalar triplets $\eta, \chi$ \footnote{In this work, the Higgs triplets
   are labeled as $\rho, \chi, \eta$ instead  of $\phi_1, \phi_2, \phi_3$ as in Ref. \cite{331b}. }containing two electrically  neutral components lying at top and bottom  of triplets: $\eta_1^0, \chi_1^0$ and  $\eta_3^0, \chi_3^0$. In the previous version \cite{r331}, only $\eta_1^0$  and $\chi_3^0$ have  vacuum expectation values (VEVs). However,
   in the new version, the $\eta_3^0$ carrying lepton number 2 has larger VEV
 of new physics scale. This leads to the mixings in both   charged and neutral gauge boson sectors. In the neutral
 gauge boson sector, the mass mixing matrix is $4 \times 4$.  In general, the diagonalizing process for $4 \times 4$ matrix
  is approximate only. However, in this paper, we show that the matrix has two exact eigenvalues and eigenstates. As a result,   the diagonalization is exact!

This  paper is organized as follows.
In Sect.\ref{model},  we briefly give particle content of the model. Sect.\ref{gauge} is devoted for gauge
boson sector. Mass mixing  matrices for charged and neutral gauge bosons are presented. The
exact solutions of $4 \times 4$ with some special feature are presented. In Sect.\ref{rho} we present the $\rho$
 parameter of the model and the equality of two VEVs at the second step of spontaneous symmetry breaking.
 We give the conclusions in the last section - Sect.\ref{conc}.

\section{\label{model}The model}
As usual \cite{r331},  the left-handed leptons are assigned
to the triplet representation of $SU(3)_L$
\be
f_L^\ell=\left(
\nu_\ell\, ,
\ell^-\, ,
N_\ell^c
 \right)^T_L  \, \sim \left(1,3, -\frac 1 3, \frac 1 3\right), \ell_R \sim (1,1,-1,1)\, ,
\ee
where $\ell=1,2,3 \equiv e,\,\mu,\,\tau$.
The numbers in bracket are assignment in $SU(3)_C, SU(3)_L, U(1)_X$ and $\mathcal{L} $.

The third quark  generation  is in triplet
\bea
Q_L^3 & = & \left(
t\, ,
b\, ,
T \right)^T_L  \, \sim \left(3,3, \frac 1 3, -\frac 2 3\right), \, T_R \sim \left(3,1,\fr 2 3,-2\right),\crn
t_R & \sim &  \left(3,1,\fr 2 3,0\right), \, b_R \sim  \left(3,1,-\fr 1 3,0\right)\, .\nn
\eea
Two first quark generations are in antitriplet
\bea
Q_L^{i} & = & \left(
d_i\, ,
-u_i\, ,
D_i
 \right)^T_L  \, \sim \left(3,3^*, 0, -\frac 2 3\right),  i= 1,2,\crn
D_{i R} & \sim & \left(3,1,-\fr 1 3,2\right),\,
u_{i R}  \sim   \left(3,1,\fr 2 3, 0\right), \, d_{i R} \sim  \left(3,1,-\fr 1 3,0\right)\, .
\nn
\eea
   In addition to the new
two-component neutral fermions present in the lepton triplet
$N_L^c\equiv (N^c)_L\equiv (\nu_R)^c$ where $\psi^c = -
C\overline{\psi}^T$, ones
introduce new sequential lepton-number-carrying gauge singlets $S=\{
S_1,S_2,S_3\}$ with the following number \cite{331b}
\[
S_i \sim (1,1,0,-1). \]

With the above $\mathcal{L}$ assignment the electric charge operator is given
 in terms of the $U(1)_X$ generator $X$ and the
diagonal generators of the $SU(3)_L$ as
\be
Q=T_3 - \frac{1}{\sqrt{3}}T_8+X  \, .
\label{dt}
\ee
Note that in the electric charge operator given  in Ref.\cite{331b}, here, the sign in front of $T_8$ is
{\it opposite}, because the leptons lie in antitriplet. If so the electrically neutral gauge bosons in
 the gauge matrix below [see Eq.(\ref{matranN})]
are $A_4, A_5$ instead of  $A_6, A_7$.

In order to spontaneously break the weak gauge symmetry, ones introduce
three scalar triplets with VEVs
\bea
\chi & = & \left(
\chi^{'0}\, ,
\chi^-\, ,
\chi^{0}
 \right)^T  \, \sim \left( 1, 3, -\fr 1 3 , \fr 4 3 \right); \hs  \langle \chi \rangle
=  \left(
0\, ,
0\, ,
n_1
 \right)^T\, ,
\label{Higg1}\\
\eta & = & \left(
\eta^{0}\, ,
\eta^-\, ,
\eta'^{0} \right)^T  \, \sim \left( 1, 3, -\fr 1 3 , -\fr 2 3 \right); \hs  \langle \eta \rangle
=  \left(
k_2\, ,
0\, ,
n_2
 \right)^T\, ,
 \label{Higg2}\\
\rho & = & \left(
\rho^{+}\, ,
\rho^0\, ,
\rho^{'+}
 \right)^T  \, \sim \left( 1, 3, \fr 2 3 , -\fr 2 3 \right); \hs  \langle \rho \rangle
=  \left(
0\, ,
k_1\, ,
0
 \right)^T\, .
 \label{Higg3}\eea
With this VEVs structure, as we see below,  the simplest consistent neutrino mass, avoiding the linear 
seesaw contribution  is realized \cite{linsesaw}. Remind that $n_2$ is a VEV of the lepton number carrying scalar,
while all of other VEVs do not.

The spontaneous symmetry breaking follows the pattern
\[
SU(3)_L\otimes U(1)_X   \stackrel{n_{1,2}}\longrightarrow    SU(2)_L\otimes U(1)_Y
\stackrel{k_{1,2}}\longrightarrow   U(1)_Q\, .
\]

 The Yukawa
Lagrangian of quark sector  is as follows \cite{331b,r331}
\bea
\mathcal{L}_{\rm quarks}&=&  y^{T}\, \overline{Q_L^{3}} \,\hat{T}_R \,\chi +
y^{D}_{i j}\, \overline{Q_L^{i}} \,\hat{D}_{R}^j \,\chi^* \crn
&+& y^{u}_{i \alpha}\, \overline{Q_L^{i}} \,\hat{u}_R^\alpha \,\rho +
y^{u}_{3,\alpha}\, \overline{Q_L^{3}} \,\hat{u}_R^\alpha \,\eta \crn
&+&
y^{d}_{3,\alpha} \,\overline{Q_L^{3}} \,\hat{d}_R^\alpha \,\rho +
y^{d}_{i \alpha} \, \overline{Q_L^{i}} \,\hat{d}_R^\alpha \,\eta
 + \mathrm{H.c.}\, .
 \label{lagQ}
\eea
The VEV $n_1$ provides masses for exotic quarks, while $n_2$ causes mixing among exotic quarks $T, D_i$ and ordinary ones.

For the lepton sector, we have \cite{331b}
\begin{eqnarray}\label{lagY}
 \mathcal{L}_{\rm leptons} &=& \,
 y^{\ell}_{ij} \overline{f_L^i} \,l_R^j \,\rho +  y^A_{ij} \, \varepsilon^{a b c }\,(\bar{f}^i_{L})_a (f^j_L)^C_b
(\rho^{*})_c  +\, y^s_{ij} \,\overline{f_L^i} \, S^j \,\chi
+ \mbox{H.c.}
\end{eqnarray}
where  $i,j=1,2,3$ is the flavor index and $a, b, c =1,2,3$ is the $\mbox{SU(3)}$ index. Note that only
 ${y^A}$ is antisymmetric and $\eta$  does not couple to
leptons.  The charged  leptons get masses the same as in the 3-3-1 model with right-handed neutrino \cite{r331}.
   The  neutrino mass matrix at the tree level,
 in the basis
($\nu_L,\,N^c,\,S$) is given by \cite{331b}
\be
M_\nu =
\left(
\begin{array}{ccc}
0& m_D &0\\
& 0 &M\\
& & 0
\end{array}
\right)\, ,
\label{mnu1}
\ee
where $m_D=k_1\,y^A$,
and $M=n_1\,y^s$.
At this level, one state $\nu_1$ is massless.  The one-loop
radiative corrections, with gauge bosons in the loop, yield a calculable
Majorana mass term \cite{331b}.  Note that the radiative seesaw is
implied for the minimal version in Ref.\cite{okada}, where the scalar bilepton
is in the loop. The obtained neutrino mass matrix and the charged lepton masses
have a strong correlation leading to leptogenesis of the model.
However, in this work, we focus our attention  only in the
gauge boson sector.

\section{Gauge boson sector}
\label{gauge}

The kinetic term for the scalar fields is
\be
\mathcal{L}_{\rm Kin} =
\sum_{H=\chi,\eta,\rho}(D^\mu H)^\dagger(D_\mu H) \,.
\label{eq1}
\ee
The covariant derivative  is
\be
D_\mu = \partial_\mu  -   i g  \, A_{a \mu } T_a -  i g'  X B_\mu T_9 ,
\label{eq2}
\ee
where $X$ is the $U(1)_X$ charge of the  field, $A_{a \mu }$ and
$B_\mu$  are  the gauge bosons  of $SU(3)_L$  and $U(1)_X$, respectively.
The above equation applies for triplet is as follows: $T_a  \rightarrow \la_a/2 \, , T_9  \rightarrow \la_9/2 $,
 where $\lambda_i$ are the Gell-Mann
matrices, and $\la_9 = \sqrt{\fr 2 3} \, \textrm{diag}\,  (1,1,1)$. The matrix    $\mathbf{A}_\mu
\equiv \sum_a A_\mu^a \lambda_a$  is
\bea
&& \mathbf{A^\mu}
=\left(%
\begin{array}{cccc}
A^\mu_3+\frac{1}{\sqrt{3}} A^\mu_8 &\sqrt{2} W_{12}^{\mu +} & A^\mu_4 - i A^\mu_5 \\
\sqrt{2} W_{12}^{\mu -} & -A^\mu_3+\frac{1}{\sqrt{3}} A^\mu_8  & \sqrt{2}W_{67}^{\mu -}\\
A^\mu_4 + i A^\mu_5&\sqrt{2} W_{67}^{\mu+} &-\frac{2}{\sqrt{3}} A^\mu_8 \\
\end{array}\,
\right). \label{matranN}
\eea

 The charged states are defined as
\be
W_{12}^{\mu \pm}=\frac{1}{\sqrt{2}} (A^\mu_1\mp i A^\mu_2)\,,\hs
W_{67}^{\mu \pm}=\frac{1}{\sqrt{2}} (A^\mu_6\pm i A^\mu_7)\,.
\label{eq3}
\ee

The mass Lagrangian of gauge fields is given by
\be
\mathcal{L}_{\rm mass} =
\sum_{H=\chi,\eta,\rho}(D^\mu \langle H \rangle)^\dagger(D_\mu \langle H \rangle) \, .
\label{eq4}
\ee
In the charged gauge boson sector, the   mass Lagrangian in  (\ref{eq4}) gives one decoupled
$A^\mu_5$ with mass \be
m^2_{A_5} = \fr{g^2}{4}(n_1^2 + n_2^2 +k_2^2)\, ,
\label{tot1}\ee
and two others with
the  mass matrix   given  in the basis of ($W^\mu_{12}, W^\mu_{67}$) as
\be
M_{charged} =
\fr{g^2}{2}\left(
\begin{array}{cc}
k_1^2 + k_2^2& n_2 k_2\\
n_2 k_2& n_1^2 + n_2^2 + k_1^2\end{array}
\right)\, .
\label{macharged}
\ee
The   matrix in Eq. (\ref{macharged}) has two eigenvalues
\be  \la_{1,2} -k^2_1 = \fr{1}{2} \left(n_1^2 + n_2^2  + k_2^2 \pm \sqrt{\De}\right)\, ,
\label{eq5}
\ee
where \bea \De  &=& (n_1^2 + n_2^2 - k_2^2)^2 + 4 n_2^2 k_2^2\
\crn
 &=& (n_1^2 + n_2^2)^2\left\{ 1+\fr{k^2_2}{(n_1^2 + n_2^2)^2}\left[2(n_2^2 - n_1^2) + k_2^2\right] \right\}\, .
\label{eqde1}
\eea
In the limit $n_1 \sim n_2 \gg k_1 \sim k_2$, one has
 \bea \sqrt{\De}
   & = & n_1^2 + n_2^2 + k_2^2  - \fr{2 n_1^2 k_2^2}{n_1^2 + n_2^2} +
 \fr{k^4_2 n_1^2}{(n_1^2 + n_2^2)^2}  \crn
  &-& \fr{k^6_2}{2(n_1^2 + n_2^2)^2}\left[1- \fr{2 n_1^2}{  n_1^2 +n_2^2}\right] + {\cal O} (k^8)
\label{eqde2}
\eea

 We will identify the light eigenvalue with square mass of the SM $W$ boson, while the heavy
 one with that of the new charged  gauge boson $Y$ carrying lepton number 2 (bilepton):
\bea m_W^2 &=& \fr{g^2}{2} \la_1 =  \fr{g^2}{2}\left[ k_1^2 +  \fr{n_1^2k_2^2}{(n_1^1+n_2^2)}  -
 \fr{n_1^2k_2^4}{(n_1^1+n_2^2)^2} \right]    + {\cal O} (k^6)
\label{eq5a},\\
m_Y^2 &=& \fr{g^2}{2} \la_2 = \fr{g^2}{2}\left[n_1^2 + n_2^2 + k_1^2 + k_2^2
-  \fr{n_1^2k_2^2}{(n_1^1+n_2^2)}+ \fr{n_1^2k_2^4}{(n_1^1+n_2^2)^2} \right]  + {\cal O} (k^6) \crn
&\simeq
 &  \fr{g^2}{2}(n_1^2 + n_2^2)\, .
\label{eq5b}
\eea
In the limit $n_1 \sim n_2 \gg k_1 \sim k_2 $, our result is consistent with that in \cite{331b}.

Two physical  bosons are determined as \cite{e331}
\bea W_\mu^-  &=& \cos \theta \,  W_{\mu 12}^-   -  \sin \theta \,  W_{\mu 67}^- \, ,\crn
Y_\mu^-  &=&  \sin \theta \,   W_{\mu 12}^-   + \cos \theta \,  W_{\mu67}^-,
\label{eq6}
\eea
where the $W-Y$ mixing angle $\theta$ charaterizing lepton number violation is given by
\be
\tan 2\theta  \equiv  \epsilon  \sim  \fr{2 n_2 k_2}{n_1^2 + n_2^2 - k_2^2}\, .
\label{eq6t}
\ee
We emphasize that due to $W - Y$ mixing, both the  $W$ boson of the SM and the bilepton $Y$  contribute
to the neutrinoless double beta decay \cite{Bilepton}.

Now we turn to the electrically  neutral gauge boson sector.
Four neutral fields, namely, $A^\mu_3,\,A^\mu_8,\,B^\mu, \, A^\mu_4$  mix
\bea
M^2 &= & \fr{g^2}{4}
\left(%
\begin{array}{cccc}
k_1^2 + k_2^2 &\fr{1}{\sqrt{3}}(k_2^2 - k_1^2)&-t\sqrt{\fr{2}{27}}(k_2^2+2k_1^2)  & n_2k_2\\
&\fr 1 3  [4(n_1^2+n_2^2)+(k_1^2 + k_2^2)]&M_{23}& -\fr{1}{\sqrt{3}}n_2k_2\\
 & &M_{33}& -2t\sqrt{\fr{2}{27}}n_2k_2 \\
& & & n_1^2+n_2^2+k_2^2 \\
\end{array}%
\right),
\crn
\label{eqAneut}
\eea
 where we have denoted $M_{23} \equiv  \fr{\sqrt{2}}{9} t[2(n_1^2 + n_2^2) + (2k_1^2 - k_2^2)]$,
 $M_{33} \equiv \fr{2t^2}{27}[(n_1^2+n_2^2)+(4k_1^2+k_2^2)] $
and $t$ is given by (see the last paper in Ref.  \cite{r331})
\be t = \fr{g'}{g} = \fr{3\sqrt{2}\sin \theta_W (m_Z')}{\sqrt{3 - 4 \sin^2 \theta_W (m_Z')} }\, .
\label{tx264}
\ee

For the matrix in  (\ref{eqAneut}), using the   programming \textsf{Mathematica}9,
 we get two \emph{exact} eigenvalues, namely,
one {\it massless} state
\[
A_\mu =  \fr{1}{\sqrt{18+4t^2}}\left( \sqrt{3}t A_{3\mu} - t A_{8\mu} + 3\sqrt{2} B_\mu
 \right)\, ,
\]
which is identified to the photon; and
the second eigenvalue  defined with
\be
m^2_{A'_4} = \fr{g^2}{2}(n_1^2 + n_2^2 +k_2^2),
\label{tot2}
\ee
associated with the eigenstate
\be
A'_{4 \mu}  =  \fr{n_2 k_2}{n_1^2 +n_2^2 -k_2^2 }A_{3 \mu} + \fr{\sqrt{3}n_2 k_2}{n_1^2
+n_2^2 -k_2^2 }A_{8 \mu} + A_{4 \mu}.
\label{tot8}\ee
In a normalized form,  the state $A'_{4 \mu}$ is rewritten as
\be
A'_{4 \mu} =
\fr{t_{2\theta}}{\sqrt{1+4t^2_{2\theta}}}A_{3\mu}+
\fr{\sqrt{3}t_{2\theta}}{\sqrt{1+4t^2_{2\theta}}}A_{8\mu}
+\fr{1}{\sqrt{1+4t^2_{2\theta}}}A_{4\mu}\, ,
\label{tot5}\ee where
$t_{2\theta}\equiv\tan2\theta$. It is emphasized that,  here the angle $\theta$  has    \emph{the same value}
  as in the charged gauge boson sector
given  in (\ref{eq6t}).

Comparing (\ref{tot1}) with (\ref{tot2}) we see that two components of $W_{45}$ have,
as  expected,  the same mass.
 Hence  we can identify
\be
X^0_\mu = \fr{1}{\sqrt{2}} (A'_{4\mu} -i A_{5\mu} )
\label{tot3}
\ee
as physical electrically  neutral non-Hermitian gauge boson. It is easy to see that this
gauge boson $X^0_\mu$ carries  lepton number 2, hence it is called bilepton gauge boson.

The  programming \textsf{Mathematica}9  also gives us two  masses of heavy physical bosons:
\bea
m_{Z_1}^2  &= &\fr{g^2}{2} \fr{1}{27}\left[ (n_1^2 + n_2^2)(18 + t^2) + k^2_2(18+4t^2)
+ k^2_1(18+t^2)- \sqrt{ \Delta'} \right]\, ,\label{klz1}\label{buoc34}\\
m^2_{Z_2} &= & \fr{g^2}{2} \fr{1}{27}\left[ (n_1^2 + n_2^2)(18 + t^2) + k^2_2(18+4t^2)
+ k^2_1(18+t^2)+  \sqrt{ \Delta'} \right].
\label{klz2}
\eea
where
\bea
 \Delta' & = &  [(n_1^2 + n_2^2 +k_2^2)(18+t^2) + 2k_1^2(9+2t^2)]^2 -
  108(9+2t^2)\left[n_1^2k_2^2 + (n_1^2 + n_2^2 +k_2^2)k_1^2\right]\crn
 & = & (n_1^2 + n_2^2)^2(18+t^2)^2\left\{ 1 + \fr{2 k_2^2}{(n_1^2 + n_2^2)}
 + \fr{4(9+2t^2)k_1^2}{(n_1^2 + n_2^2)(18+t^2)}\right. \crn
 & - &  108 \fr{(9+2t^2)}{(n_1^2 + n_2^2) (18+t^2)^2} \left[ k_1^2
 + \fr{n_1^2 k_2^2 }{(n_1^2 + n_2^2)}\right]\crn
 & + &\left. \fr{1}{(n_1^2 + n_2^2)^2} \left[ k_2^4+ \fr{4(9+2 t^2)^2 k_1^4}{(18+t^2)^2} +
   \fr{4(9+2 t^2)(t^2 -9) k_1^2k_2^2}{(18+t^2)^2}\right]\right\}\, .
\label{buoc35t}
\eea
%Da kiem tra 26.4-------
Then
\bea
\sqrt{ \Delta'} & = &   (n_1^2 + n_2^2)(18+t^2)
 + k_2^2  (18+t^2) +  2(9+2t^2)k_1^2
  -   54 \fr{(9+2t^2)}{ (18+t^2)} \left( k_1^2 + \fr{n_1^2 k_2^2 }{n_1^2 + n_2^2}\right)\crn
 &+&\fr{(18+t^2)}{(n_1^2 + n_2^2)}\left\{ \fr{54(9+2t^2)}{ (18+t^2)^4} \left( k_1^2
 + \fr{n_1^2 k_2^2 }{n_1^2 + n_2^2}\right)\left[k_2^2(18+t^2)^2 + 2k_1^2(18+t^2)(9+2t^2)
\right. \right.  \crn& -&\left. \left.  54(9+2t^2)\left( k_1^2
+ \fr{n_1^2 k_2^2 }{n_1^2 + n_2^2}\right)\right]-\fr{56(9+2t^2)k_1^2 k_2^2}{(18+t^2)}\right\}
+ \mathcal{O}(k^6) \, .
\label{buoc35t2}
\eea
%Da kiem tra 26.4-------
Substituting (\ref{buoc35t2}) into (\ref{buoc34}) yields the mass of the  physical $Z_1$ boson:
\bea
m_{Z_1}^2 & =  & \fr{g^2}{54}\left\{  \fr{54(9+2t^2)}{18+t^2}\left( k_1^2 +
\fr{n_1^2 k_2^2}{ (n_1^2 + n_2^2)} \right)  + 3 t^2(k_2^2-k_1^2) \right.
\crn &-&  \fr{(18+t^2)}{(n_1^2 + n_2^2)}\left\{ \fr{54(9+2t^2)}{ (18+t^2)^4}
\left( k_1^2 + \fr{n_1^2 k_2^2 }{n_1^2 + n_2^2}\right)\left[k_2^2(18+t^2)^2 +  2k_1^2(18+t^2)(9+2t^2)
\right. \right.  \crn& -&\left. \left. 54(9+2t^2)\left( k_1^2
+ \fr{n_1^2 k_2^2 }{n_1^2 + n_2^2}\right)\right]-\fr{56(9+2t^2)k_1^2 k_2^2}{(18+t^2)}\right\}
+ \mathcal{O}(k^6)\, .
\label{masZ1}
\eea
%Da kiem tra 26.4-------
It is emphasized that, at the leading order,  there are two terms (in first line of Eq.(\ref{masZ1})):
one is the main mass term of the $Z$ boson
and the second one is the unusual  difference of square VEVs: $(k_1^2 - k_2^2)$. This term  leads to
an interesting equality below.

Similarly,  for the physical heavy extra  neutral gauge boson $Z_2$, one obtains
\bea
m_{Z_2}^2  & =  & \fr{g^2}{27} \left\{ ( n_1^2 + n_2^2 +k_1^2+k_2^2)(18+t^2) +\fr 3 2 t^2 (k_1^2  +k_2^2)
-  \fr{27(9+2t^2)}{18+t^2}\left( k_1^2 +
\fr{n_1^2 k_2^2}{ (n_1^2 + n_2^2)} \right)   \right.
\crn &+&  \fr{(18+t^2)}{(n_1^2 + n_2^2)}\left( \fr{27(9+2t^2)}{ (18+t^2)^4}
\left( k_1^2 + \fr{n_1^2 k_2^2 }{n_1^2 + n_2^2}\right)\left[k_2^2(18+t^2)^2 + 2k_1^2(18+t^2)(9+2t^2)
\right. \right.  \crn& -&\left. \left. \left.  54(9+2t^2)\left( k_1^2
+ \fr{n_1^2 k_2^2 }{n_1^2 + n_2^2}\right)\right]-\fr{28(9+2t^2)k_1^2 k_2^2}{(18+t^2)}\right)\right\}
+ \mathcal{O}(k^6)\, .\crn
&\simeq & \fr{g^2( n_1^2 + n_2^2)(18+t^2) }{27} \, .
\label{buoc35t4}
\eea
%Da kiem tra 26.4-------

Due to the quark family
discrimination in the model,  $Z'$/$Z_2$ couples  nonuniversally  to  the ordinary quarks, it gives rise to tree-level
flavour-changing neutral current (FCNC) \cite{fncn}. This would induce gauge-mediated FCNCs, e.g.,
$b \rightarrow s \mu^+ \mu^-$  \cite{buras}, providing a test of the model.
We finish this section by remark that the gauge boson mixing here is completely  similar to that
of the economical 3-3-1 model (ECN331) \cite{e331}.  However, the key difference is that,
 here the lepton number carrying
VEV $n_2$ is very large $(n_2 \sim n_1 \gg k_1 \sim k_2)$,  while in the ECN331 model, the  lepton number carrying
VEV $u$ is very small $(u \ll v)$ with $v \simeq 245$ GeV. Within our result, in the  figure 1  of  Ref.\cite{331b},
 the unphysical gauge field $W_6$ is replaced by {\it physical field}  $X^0$, while $W^3, W^8, B$ are replaced by
 physical neutral gauge bosons   $Z_1, Z_2$.
 However,  the result is  the same.

The diagonalization process of the mass matrix of neutral gauge bosons, the currents and the
model phenomenology will be analyzed in details elsewhere.

\section{Michel   parameter $\rho$}
\label{rho}
As seen from above, the unusual term in the   $Z_1$ boson mass will affect
the well-determined parameter $\rho$. Thus,
for our purpose we consider the $\rho$ parameter - one of the most
important quantities of the SM, having a leading contribution in
terms of the $T$ parameter
\be \rho = 1 + \al T\, . \label{rhot} \ee
 In the usual
3-3-1 model, $T$ gets contribution from the $Z-Z'$ mixing  and the oblique correction
 \cite{il} \[ T = T_{Z Z'} +
T_{oblique},\] where $ T_{Z Z'} \simeq \fr{\tan^2
\varphi}{\al}\left( \fr{m^2_{Z_2}}{m^2_{Z_1}} -1 \right)$ is
negligible for $m_{Z'}$  less than 1 TeV,  $T_{oblique}$ depends on
masses of  the top quark and the SM  Higgs boson.

At  the tree level, from (\ref{eq5a}) and
(\ref{masZ1}) we get an expression for the $\rho$ parameter in
the  model under consideration
 \bea \rho & = & \fr{m^2_{W}}{c^2_w
m^2_{Z_1}}  =  \fr{18+t^2}{2(9+2 t^2)c^2_w}\left\{ 1 + \fr{t^2(18+t^2)(k_1^2-k_2^2)}{18(9+2t^2)
\left( k_1^2 + \fr{n_1^2 k_2^2 }{n_1^2 + n_2^2}\right)} \right.
\crn &-&\fr{n_1^2k_2^4}{(n_1^2+n_2^2)^2\left( k_1^2 + \fr{n_1^2 k_2^2 }{n_1^2 + n_2^2}\right)}
 +
  \fr{1}{(n_1^2 + n_2^2)(18+t^2)^2}\left(\left[k_2^2(18+t^2)^2 + 2k_1^2(18+t^2)(9+2t^2)
 \right. \right.  \crn& - &
  \left. \left. \left. 54(9+2t^2)\left( k_1^2 + \fr{n_1^2 k_2^2 }{n_1^2 + n_2^2}\right)\right]-\fr{56k_1^2 k_2^2}{54
  \left( k_1^2 + \fr{n_1^2 k_2^2 }{n_1^2 + n_2^2}\right)}\right)\right\}+ \mathcal{O}(k^6)\ \, ,
\label{rhopa421} \eea
%Da kiem tra 26.4-------
where we have denoted $s_w \equiv \sin \theta_w, c_w \equiv \cos \theta_w, t_w \equiv \tan \theta_w$, and
so forth.
Two terms in the first line of Eq. (\ref{rhopa421}) do not depend on perturbative  small value ($k/n$),
 where  $ k \approx k_1, k_2\, , n \approx n_1, n_2$;  and
they are the leading order of the $\rho$ parameter.

Experimental data \cite{pdg}  show that the $\rho$ parameter is very close with the unit
\be
\rho = 1.01031 \pm 0.00011 \, .
\label{rhopart16}
\ee
Hence,  at the leading order, the following  requirement should be fulfilled
\be
 \fr{(18+t^2)}{2(9+2 t^2)c^2_w}\left[ 1 + \fr{t^2(18+t^2)(k_1^2-k_2^2)}{18(9+2t^2)
 \left( k_1^2 + \fr{n_1^2 k_2^2 }{n_1^2 + n_2^2}\right)} \right]=1\, .
\label{rhopa421t} \ee
Substituting (\ref{tx264}) into  (\ref{rhopa421t})  yields
\be
  \fr{2c_w^2s_w^2}{(3-4s_w^2)} \fr{(k_1^2-k_2^2)}{\left[k_1^2 +
\fr{n_1^2k_2^2}{(n_1^2+n_2^2)}\right]} = 0\, .
\label{rhopart} \ee
 Thus,   we obtain the relation
\be k_1 = k_2\, .
\label{hay}
\ee
Note that, {\it in the first time,} the equality in (\ref{hay}) exists  in the model under consideration.
This will helpful in our future study. To get constraint from $\rho$ parameter, we should
include oblique corrections; and for more details, the reader is referred to \cite{dongsi}.

\section{Conclusion}
\label{conc}
In this paper, we have showed  that the mass matrix of electrically neutral  gauge bosons in the recently
 proposed  3-3-1 model
with three lepton/neutrino singlets \cite{331b}  has two exact eigenvalues and corresponding eigenvectors.
With two determined eigenvalues,  the $4\times 4$ mass matrix
is diagonalized exactly. Two components of neutral bilepton boson $X^0_\mu$ have the same mass,
hence the neutral non-Hermitian gauge boson $X^0_\mu$ is properly determined. 
%This contradicts to previous analysis in Ref.\cite{331b}.
With extra vacuum expectation values of the Higgs fields, there are mixings among charged gauge
bosons $W^\pm$ and $Y^\pm$ as well as among
neutral gauge bosons $Z, Z'$ and $X^0$. Due to these mixings, the lepton number violating
interactions exist in leptonic  currents
  not only in bileptons  $Y$ and $X^0$ but also in  both SM  $W$ and $Z$ bosons.
The mixing of gauge bosons in the model under consideration leads to some anomalous couplings
of both $W$ and $Z$ bosons, which are subject of our next works.

 The scale of new physics
was estimated to be in range of few TeVs. With this limit, masses of the exotic quarks are also not high, in the range
of few TeVs. The leading order of the Michel parameter requires the equality: $k_1=k_2$,  which is obtained
 in the first time. The derived relation  will ease our future study.
The above mentioned mixings lead to new anomalous currents and
  very rich phenomenology.  The model is interesting and  deserves  further intensive studies.

 \section*{Acknowledgment}
I thank Phung Van Dong for  consultation  in Mathematica and useful  remarks.  This research is funded by Vietnam  National Foundation for Science and Technology Development (NAFOSTED)  under grant number
103.01-2014.51.
\\[0.3cm]

\end{document}